\documentclass[preprint]{revtex4}
\usepackage{graphicx,float}
\usepackage{amssymb,amsmath}

\begin{document}
\title{Effects of dark matter pressure on the ellipticity of cosmic voids}

\author{Zeinab Rezaei\footnote{E-mail: zrezaei@shirazu.ac.ir}}

\affiliation{Department of Physics, Shiraz
University, Shiraz 71454, Iran.\\
Biruni Observatory, Shiraz
University, Shiraz 71454, Iran.}

\begin{abstract}
The dark matter in or around the cosmic voids affects their shapes. The thermodynamical properties of
dark matter can alter the ellipticity of cosmic voids. Here, applying the dark matter equation of state from the pseudo-isothermal density profile of galaxies, we explore the shapes of cosmic voids with the non zero pressure dark matter in different cosmological models. For this purpose,
the linear growth of density perturbation in the presence of dark matter pressure is calculated.
In addition, the matter transfer function considering the dark matter pressure, as well as the linear matter power spectrum in the presence of the dark matter pressure are presented. Employing these results, the probability density distribution for the ellipticity of cosmic voids with the non zero pressure dark matter is calculated. Our calculations verify that the dark matter pressure leads to more spherical shapes for the cosmic voids.

\end{abstract}
\maketitle
\section{Introduction}
\label{intro}
In cosmic web, the large empty regions with very low number density of galaxies are called cosmic voids. The cosmic voids as the bubbles of the universe, are separated by the sheets and filaments in which the clusters and superclusters are observed.
To investigate  the large-scale structure of the Universe, it is important to study the properties of the cosmic voids.
Considering the standard gravitational instability theory, the voids form from the local minima of the initial
density field and expand faster than the rest of the Universe \cite{Park}. Because the dark energy (DE) and dark matter (DM) can affect the properties of cosmic voids, they are counted as cosmological
probes.
In this regard, the shape of the cosmic void is considered as an observable. The DM around the cosmic voids has tidal effects which this leads to the distortion of the void shape. Therefore, the susceptibility of
the void shapes to the tidal distortions can be an indicator of the large-scale tidal and density fields \cite{Lee5}. Moreover, the background cosmology determines the shape evolution of the voids \cite{Lee9}.

The ellipticity of the voids as a result of the tidal field effects,
is an important observable related to the void shapes. Many works have been done to
investigate the ellipticity of cosmic voids \cite{Park,Lee9,Lavaux,Biswas,Bos,Ricciardelli,Leclercq,Massara,Adermann}.
The void ellipticity distribution as a result of the
counterbalance between the tidal effect of the DM and the expansion of the Universe depends on the cosmological parameters and
moves toward the low ellipticity as the smoothing scale increases \cite{Park}.
The mean ellipticity of voids decreases as the redshift grows with a rate which depends on the DE equation of state (EOS) \cite{Lee9}.
The voids with larger smooth scales become more spherical and the ellipticity distribution moves towards a perfect sphere \cite{Lavaux}.
The variance of the
fluctuations and the amplitude of primordial perturbations as well as the spectral
index affect the distribution of ellipticities \cite{Biswas}.
Studying the influence of the darkness on the void shape in five sets of cosmological N-body simulations
shows that the mean ellipticity of the voids grows with time \cite{Bos}.
Eulerian cosmological codes confirm that by increasing the redshift, the ellipticity decreases due to the influence of the tidal field of the large-scale structure \cite{Ricciardelli}.
The results related to
the ellipticity distribution of voids in non-linear gravitational model with data-constrained reconstructions
of the observed large-scale structure are in agreement with the DM simulations \cite{Leclercq}.
Using a numerical study of the statistical properties of voids in both
massive and massless neutrino cosmologies, the effects of neutrinos on the void
ellipticities has been investigated \cite{Massara}.
Applying N-body simulations with evolving and interacting
dark sectors shows that the smaller voids
exhibit a greater spread in ellipticity than the larger voids \cite{Adermann}.

Since the DM as one of the important portion of the universe can affect the voids, many
studies have considered the DM in or around the voids \cite{Ghigna4,Ghigna6,Miranda,Benson,Sheth,Gubser,Nusser,Brunino,Hahn8,Cuesta,Peebles,Kreckel,Bos,Sutter,Yang,Leclercq}.
The void statistics in a
volume-limited subsample of the Perseus-Pisces survey is sensible to the shape of the linear spectrum, the nature of the DM, and the passage from CDM to cold+hot DM \cite{Ghigna4,Ghigna6}.
In the formation of cosmological structures in the early Universe, the physical processes acting on the baryonic matter produce a transition region where the radius of the DM component is greater than the baryonic void radius \cite{Miranda}.
Applying a semi-analytic model of galaxy formation verifies that the voids in the distributions of galaxies and DM have different statistical properties \cite{Benson}.
Moreover, the voids in the galaxy distribution are similar
to those in the DM distribution \cite{Sheth} .
Long-range scalar forces result in the expulsion of DM halos from the voids and make the voids
more empty \cite{Gubser}.
Numerical simulations verify that a long-range scalar interaction in a single species of massive DM particles results in the lower density of DM in the voids \cite{Nusser}.
The Millennium N-body simulation for the galaxy DM haloes surrounding the largest cosmological voids confirms that
the major axes of the DM haloes lies parallel to the surface of the voids \cite{Brunino}.
The importance of the physical mass scale and the environmental role in the evolution of DM haloes in voids have been investigated \cite{Hahn8}.
The high resolution cosmological N-body simulation shows that galaxy-size DM haloes around voids have spins that lie in the shells of voids \cite{Cuesta}.
The apparently tranquil environment in voids in the distribution of large galaxies suggests the best
chance for survival of low-mass DM haloes \cite{Peebles}.
By analyzing an adaptive mesh refinement hydrodynamic simulation, it has been found that the most massive DM halos avoid the void center \cite{Kreckel}.
Comparing voids in the DM distribution to voids in the halo population shows the statistically significant sensitivity of voids in the DM distribution \cite{Bos}.
Applying halo occupation distribution models verifies that the voids which are large in galaxies are also large in DM \cite{Sutter}.
In the warm DM (WDM) model, the voids are shallower and larger and also
the void density minima grow shallower with the increase of
DM warmth \cite{Yang}.
Small galaxy voids are more elliptical than DM voids because of
important Poisson fluctuations below the mean galaxy separation \cite{Leclercq}.

Although the DM is usually considered as a pressureless fluid, some investigations have shown
that it can be a fluid with non zero pressure. Many works have been done to study the EOS of DM and
the effects of DM pressure \cite{Bharadwaj,Muller,Faber,Binder,Nakajima,Su,Saxton,Bezares,Wechakama,Guzman,Harko1,Harko2,Bettoni,Barranco,Kunz}.
The EOS of DM can be determined using the combination of observations related to rotation curves and gravitational lensing \cite{Bharadwaj}.
In a modified $\Lambda$CDM cosmology, applying CMB, supernovae Ia, and large scale structure data,
the constraints on the EOS of DM have been obtained \cite{Muller}.
The density and pressure profiles of the galactic fluid have been calculated
using the observations of galaxy rotation curves and gravitational lensing \cite{Faber}.
With non-vanishing DM pressure,
negative small values of the coupling constant result in a decelerated-accelerated transition at
lower red-shifts with a better adjustment of the present value of the deceleration parameter of the Universe \cite{Binder}.
The flat-top column density profile of clusters of galaxies
could be explained by the effects of degeneracy pressure of fermionic DM \cite{Nakajima}.
The gravitational lensing deflection angle
can determine the non-ideal fluid EOS for the DM halo using observations \cite{Su}.
For DM haloes, the polytropic EOS can explain the
extended theories of DM considering self-interaction, non-extensive thermostatistics, and
boson condensation \cite{Saxton}.
Studying the scalar-field excitations of induced gravity with a Higgs potential shows that the DM dominance leads to pressures related to an EOS parameter of total energy of the same value as for weak fields in solar-relativistic ranges \cite{Bezares}.
The pressure from DM annihilation could set constraints on the inner slope of halo density
profile and the mass and the annihilation cross-section of DM particles into
electron-positron pairs \cite{Wechakama}.
For the DM with non trivial pressure near a supermassive black hole,
the contribution of accreted DM to the supermassive black hole growth could be small \cite{Guzman}.
The possibility that DM is a mixture of two different non-interacting perfect fluids
which can be considered as an effective single anisotropic fluid with distinct radial and tangential pressures
has been studied \cite{Harko1}.
A general r-dependent functional relationship between the energy density and the
radial pressure of the DM halos exists \cite{Harko2}.
The non-minimal coupling between gravity
and DM translates into an effective pressure for the DM component and this effective pressure reduces the growth of structures at galactic scales \cite{Bettoni}.
The galactic halos describing the DM as a non zero pressure
fluid has been modeled and a DM EOS using the observational
data from the rotation curves of galaxies has been obtained \cite{Barranco}. This EOS has a functional dependence universal
for all galaxies.
Using large-scale cosmological observations, it is possible to
place constraints on the DM pressure \cite{Kunz}.
Noting the above discussions about the DM pressure, it is possible to conclude that
the DM pressure can affects the ellipticity of voids. In the present work, we
study the effects of the DM pressure on the ellipticity of voids.

\vspace{-0.58cm}
\section{Dark mater equation of state}

The DM EOS has been obtained applying the observational data of the rotational curves of galaxies \cite{Barranco}. The pseudo-isothermal model leads to a mass density profile with the property of
regularity at the origin. If one uses the velocity profile, geometric potentials, and gravitational potential,
the DM EOS given by the pseudo-isothermal density profile is as follows \cite{Barranco},
\begin{eqnarray}\label{213}
       {P_{DM}}({\rho_{DM}})=\frac{8  {p}_0}{\pi^2-8}[\frac{\pi^2}{8}-\frac{arctan\sqrt{\frac
       {{\rho}_0}{{\rho_{DM}}}-1}}{\sqrt{\frac{{\rho}_0}
       {{\rho_{DM}}}-1}}
	 -\frac{1}{2}(arctan\sqrt{\frac
       {{\rho}_0}{{\rho_{DM}}}-1}\ )^2].
 \end{eqnarray}
In the above equation, $\rho_{DM}$ and $P_{DM}$ denote the density and pressure of DM. Besides,
the free parameters, $\rho_0$ and $p_0$, are the central density and pressure of galaxies.
It has been argued that this DM EOS has a functional dependence
which is universal for all galaxies and it is possible to predict the central pressure and density of
the galaxies employing this DM EOS and the rotational curve data. Following this universality, we apply
this DM EOS for the DM in or around the voids.
%%%%%%%%%%%%%%%%%%%%%%%%%%%%%%%%%%%%%%%%%%%%%%%%%%%%%%%%%%%%%%%%%%%%%%%%%%%%%%%%%%%%%%%%%%%%%%%%%%%
\begin{figure*}
\vspace*{1cm}       % Give the correct figure height in cm
\includegraphics[width=0.36\textwidth]{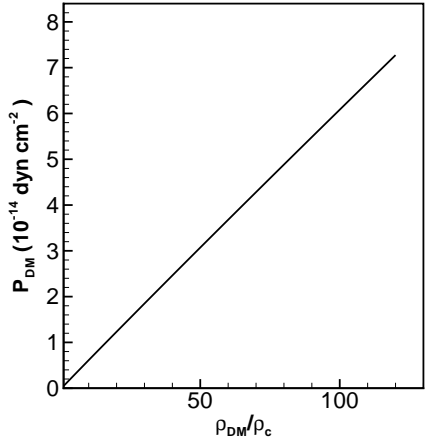}
\caption{Dark matter EOS related to the
galaxy U5750 with the parameters ${\rho}_0=0.31\ GeV/cm^3$ and
$p_0=1.1\times 10^{-8} \ GeV/cm^3$ and $\chi^2_{min}/d.o.f.=0.01$ \cite{Barranco}.}
\label{fig1}
\end{figure*}
In this work, for the DM with nonzero pressure, we use the DM EOS related to the
galaxy U5750 with the parameters ${\rho}_0=0.31\ GeV/cm^3$ and
$p_0=1.1\times 10^{-8} \ GeV/cm^3$ \cite{Barranco}. This DM EOS which is the result of one of the best fit with
$\chi^2_{min}/d.o.f.=0.01$ has been shown in Figure \ref{fig1}.
%%%%%%%%%%%%%%%%%%%%%%%%%%%%%%%%%%%%%%%%%%%%%%%%%%%%%%%%%%%%%%%%%%%%%%%%%%%%%%%%%%%%%%%%%%%%%%%%%%%%

\vspace{-0.58cm}
\section{Linear growth of density perturbation in the presence of dark matter pressure}

\begin{figure*}
\vspace*{1cm}       % Give the correct figure height in cm
\includegraphics[width=1.0\textwidth]{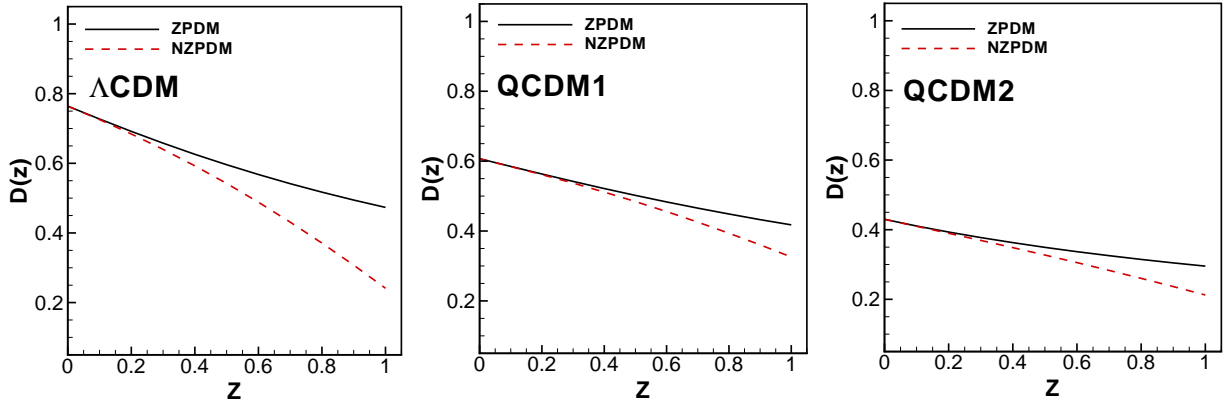}
\caption{Linear growth factor, $D(z)$, as a function of the redshift, $z$,
for two cases of zero pressure DM (ZPDM) and non zero pressure DM (NZPDM),
applying different cosmological models.}
\label{fig4}
\end{figure*}

In 1957, Bonnor obtained the differential equation describing the growth of the matter density perturbation, $D(t)=\delta\rho_{M}/\rho_{M}$, for a cloud of gas with the EOS $P_{M}(\rho_{M})$ as follows \cite{Bonnor},
\begin{eqnarray}\label{Bonnor1}
       a^2 \ddot{D}+2a\dot{a}\dot{D}+(N^2\frac{dP_{M}}{d\rho_{M}}-4\pi G \rho_{M0}a^2)D=0,
 \end{eqnarray}
in which $a(t)$ is the cosmic scale factor, $\rho_{M}$ is the matter density, $\rho_{M0}$ is the matter density at the present time, and $N$ is a real constant. Moreover, $\rho_{M}=\rho_{DM}+\rho_{VM}$ and $P_{M}=P_{DM}+P_{VM}$, in which DM and VM denote the dark and visible matter, respectively. In Eq. (\ref{Bonnor1}), the value of the $\frac{dP_{M}}{d\rho_{M}}$ is calculated at the present time \cite{Bonnor}. Besides, the cosmic scale factor, $a(t)$, satisfies the Friedmann equations (using the units in which $c=1$),
\begin{eqnarray}\label{f1}
\dot{a}^2+k =\frac{8\pi G}{3}\rho a^2,
 \end{eqnarray}
\begin{eqnarray}\label{f2}
\dot{a}^2+k +2a \ddot{a} =-8\pi G P a^2.
 \end{eqnarray}
In the above equations, the total energy density, $\rho$, is related to the matter
energy density, $\rho_{M}$, radiation energy density, $\rho_{R}$, and DE density, $\rho_{DE}$, by $\rho=\rho_{M}+\rho_{R}+\rho_{DE}$. In addition, the total pressure, $P$,
is expressed in terms of the matter pressure, $P_{M}$, radiation pressure, $P_{R}$, and DE pressure, $P_{DE}$, as $P=P_{M}+P_{R}+P_{DE}$. Besides, $k=-1,0$, and $1$ for an open, flat, and closed universe, respectively. It has been shown that neglecting the matter pressure, i.e. $P_{M}=0$, and considering the radiation EOS $P_{R}=\frac{1}{3}\rho_{R}$ as well as the DE EOS $P_{DE}= w(a) \rho_{DE}$ in which $w(a)=w_0+w_a(1-a)$, the solution to Eq. (\ref{Bonnor1}) is \cite{Heath,Percival,Lee9},
\begin{eqnarray}\label{Heath1}
D(a)=\frac{5\Omega_M}{2}E(a)\int_0^a\frac{da'}{[a'E(a')]^3},
 \end{eqnarray}
with the matter density parameter at the present time, $\Omega_M$, and
\begin{eqnarray}\label{Heath2}
E(a)=[\Omega_M a^{-3}+\Omega_R a^{-4}+\Omega_{DE}\ a^{f(a)}]^{1/2}.
 \end{eqnarray}
In Eq. (\ref{Heath2}), $\Omega_{R}$ and $\Omega_{DE}=1-\Omega_M-\Omega_{R}$ are the radiation and DE density parameter and the function $f(a)$ is given by,
\begin{eqnarray}\label{Heath3}
f(a)=-3(1+w_0)+\frac{3w_a}{2lna}.
 \end{eqnarray}
The result of the Eq. (\ref{Heath1}) for the linear growth factor $D(z)$ versus the redshift $z$ which is related to the
scale factor by $a=(1+z)^{-1}$ has been presented in Figure \ref{fig4} with the solid curves.
It should be noted that we have applied three cosmological models, i.e. $\Lambda$CDM, QCDM1, and QCDM2. For all these models $w_a=0.0$, while $w_0=-1.0,-2/3,$ and $-1/3$  for $\Lambda$CDM, QCDM1, and QCDM2, respectively. Besides, the value of the cosmological parameters are $\Omega_M=0.276$ and $\Omega_{R}=8.6\times10^{-5}$.

However, in this work, we are interested in the effects of the DM pressure on the linear growth factor
as well as the ellipticity of voids. Therefore, considering $P_{M}=P_{DM}$ as the matter EOS and applying this EOS to
Eqs. (\ref{Bonnor1})-(\ref{f2}), we calculate the redshift dependency of the linear growth factor.
To do this we first rewrite Eq. (\ref{Bonnor1}) in terms of the derivative of $D(z)$ respect to the redshift, $z$,
\begin{eqnarray}\label{Bonnor2}
&&a^{-2} \dot{a}^2 \frac{d^2 D(z)}{dz^2}-\ddot{a}\frac{d D(z)}{dz}+(N^2\frac{dP_{DM}}{d\rho_{DM}}-4\pi G \rho_{M0}a^2)D(z)=0,
 \end{eqnarray}
in which $P_{DM}$ is given by Eq. (\ref{213}). Besides, it is easy to show that Eqs. (\ref{f1})-(\ref{f2}) leads to the following relations,
\begin{eqnarray}\label{ff1}
\dot{a}^2 =\frac{8\pi G}{3}\rho a^2-k ,
 \end{eqnarray}
\begin{eqnarray}\label{ff2}
\ddot{a}=-4\pi G(P+\frac{\rho}{3})a.
 \end{eqnarray}
Substituting  Eqs. (\ref{ff1})-(\ref{ff2}) into Eq. (\ref{Bonnor2}) and multiply both sides by $a^{-2}$ give,
\begin{eqnarray}\label{Bonnor3}
&&(\frac{8\pi G}{3}\rho a^{-2} -k a^{-4}) \frac{d^2 D(z)}{dz^2}+4\pi G(P+\frac{\rho}{3})a^{-1}\nonumber \\&\times&\frac{d D(z)}{dz}+(N^2\frac{dP_{DM}}{d\rho_{DM}}a^{-2}-4\pi G \rho_{M0})D(z)=0.
 \end{eqnarray}
Noting the equality $a^{-1}=1+z$ and considering the case $k=0$, the above equation leads to the following Bessel equation,
\begin{eqnarray}\label{Bonnor4}
e_1(1+z)^2\frac{d^2 D(z)}{dz^2}+e_2(1+z)\frac{d D(z)}{dz}+[e_3(1+z)^2-e_4]D(z)=0,
 \end{eqnarray}
with
\begin{eqnarray}\label{Bonnor5}
e_1=\frac{8\pi G}{3}\rho,
 \end{eqnarray}
\begin{eqnarray}\label{Bonnor6}
e_2=4\pi G(P+\frac{\rho}{3}),
 \end{eqnarray}
\begin{eqnarray}\label{Bonnor7}
e_3=N^2\frac{dP_{DM}}{d\rho_{DM}},
 \end{eqnarray}
\begin{eqnarray}\label{Bonnor8}
e_4=4\pi G \rho_{M0}.
 \end{eqnarray}
In Eqs. (\ref{Bonnor5})-(\ref{Bonnor8}), the values of $e_i$ are calculated at the present time.
Moreover, introducing the wave-length $\lambda$ of the disturbance by $\lambda=2\pi a/N$ (in which $a=1$ at the present time), it has been argued that the
disturbance increases exponentially with time if
\begin{eqnarray}\label{Bonnor9}
\lambda >(\frac{\pi}{G\rho_{M0}}\frac{dP_{DM}}{d\rho_{DM}}|_{\rho_{DM0}})^{1/2},
 \end{eqnarray}
\cite{Bonnor}. Therefore, in  Eq. (\ref{Bonnor7}), $N$ should satisfy
\begin{eqnarray}\label{Bonnor10}
N^2<\frac{4\pi G \rho_{M0}}{\frac{dP_{DM}}{d\rho_{DM}}|_{\rho_{DM0}}}.
 \end{eqnarray}
Considering the maximum value of $N$, Eq. (\ref{Bonnor7}) leads to
\begin{eqnarray}\label{Bonnor11}
e_3=4\pi G \rho_{M0}.
 \end{eqnarray}
The solution to Eq. (\ref{Bonnor4}) is
\begin{eqnarray}\label{Bonnor12}
D(z)=[c_1J_{\alpha}(\sqrt{\frac{e_3}{e_1}}(1+z))+c_2N_{\alpha}(\sqrt{\frac{e_3}{e_1}}(1+z))]\times(1+z)^{\xi}.
 \end{eqnarray}
In the above relation, $J_{\alpha}$ and $N_{\alpha}$ are the Bessel functions of the first and second kinds, respectively.
Besides, the parameters $\alpha$ and $\xi$ are given by the following relations,
\begin{eqnarray}\label{Bonnor13}
\alpha=\frac{1}{2e_1}\sqrt{e_1^2+e_2^2+(4e_4-2e_2)e_1},
 \end{eqnarray}
\begin{eqnarray}\label{Bonnor14}
\xi=\frac{e_1-e_2}{2e_1}.
 \end{eqnarray}
Moreover, the constant values $c_1$ and $c_2$ in Eq. (\ref{Bonnor12}) should be determined through the boundary conditions.
In this regard, we consider the following boundary conditions,
\begin{eqnarray}
D_{P_{DM}\neq0}(z=0)=D_{P_{DM}=0}(z=0),
 \end{eqnarray}
 and
\begin{eqnarray}
\frac{dD_{P_{DM}\neq0}}{dz}|_{z=0}=\frac{dD_{P_{DM}=0}}{dz}|_{z=0}.
 \end{eqnarray}
In fact, we assume that the linear growth of density perturbation and its derivative respect to z are equal for
zero pressure DM and non zero pressure DM at the present time.

The results for the linear growth of density perturbation taking the DM pressure into account in different cosmological models have been presented in Figure \ref{fig4} with the dashed curves.
For each cosmological model and at any value of the redshift, the linear growth factor
is smaller in the case with non zero pressure DM (NZPDM) compared to the case of zero pressure DM (ZPDM).
The effects of DM pressure on the linear growth factor is more significant at higher values of
the redshift. Besides, the DM pressure results in the increase of
the rate at which the linear growth factor decreases with the redshift.
Figure \ref{fig4} confirms that the influence of the DM pressure on the linear growth factor
is more considerable in the $\Lambda$CDM model compared to QCDM1 and QCDM2 models.

\vspace{-0.58cm}
\section{Matter transfer function in the presence of the dark matter pressure}

\begin{figure*}
\vspace*{1cm}       % Give the correct figure height in cm
\includegraphics[width=.4\textwidth]{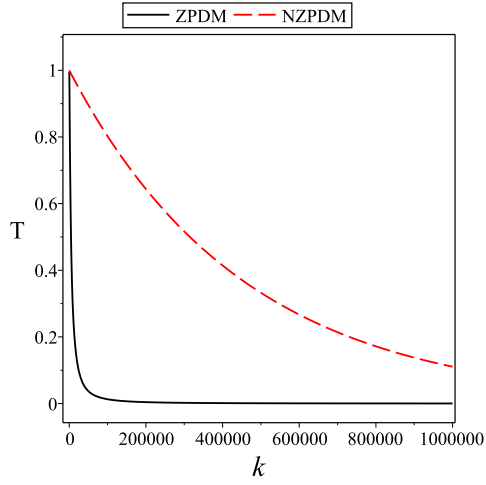}
\caption{Matter transfer function in the cases of ZPDM and NZPDM.}
\label{fig6}
\end{figure*}

In the linear perturbation theory, the measuring of the perturbations at large scales
is considered by comparing them to the amplitude they would
have had neglecting causal physics \cite{Eisenstein}. This behavior is described by the
transfer function which is given by
\begin{eqnarray}\label{Trans}
T(k)\equiv\frac{\delta(k,z=0)\delta(0,z=\infty)}{\delta(k,z=\infty)\delta(0,z=0)},
 \end{eqnarray}
in which $\delta(k,z)$ denotes the density perturbation for wavenumber $k$
and redshift $z$.
Considering the Einstein equations and the energy momentum conservation, the transfer function in the case of ZPDM can be obtained as follows \cite{Weinberg,Chongchitnan},
\begin{eqnarray}\label{Tzpdm}
T_{ZPDM}(x)=\frac{ln[1 + (0.124x)^2]}{(0.124x)^2}\times (\frac{1 + (1.257x)^2 + (0.4452x)^4 + (0.2197x)^6}{1 + (1.606x)^2 + (0.8568x)^4 + (0.3927x)^6})^{1/2},
 \end{eqnarray}
and
\begin{eqnarray}
x_{EH}=\frac{k\Omega_R^{1/2}}{H_0\Omega_M}[\alpha+\frac{1 -\alpha}{1 + (0.43ks)^4}]^{-1},
 \end{eqnarray}
where
\begin{eqnarray}
\alpha=1-0.328 ln(431\Omega_M h^2)\frac{\Omega_B}{\Omega_M}+ 0.38 ln(22.3\Omega_M h^2)(\frac{\Omega_B}{\Omega_M})^2,
 \end{eqnarray}
 and
 \begin{eqnarray}
s =\frac{44.5 ln(9.83/\Omega_M h^2)}{\sqrt{1 + 10(\Omega_B h^2)^{3/4}}} Mpc.
 \end{eqnarray}
It should be noted that in the above equations, $H_0\equiv100h\ km\ s^{-1}\ Mpc^{-1}$ is the Hubble constant and
$\Omega_B$ with the value $\Omega_B=0.046$ denotes the density parameter for the baryons, only.
Besides, we have applied the value $h=0.704$.
Figure \ref{fig6} presents the matter transfer function for ZPDM with the solid curve. However, here we are interested in determining the matter transfer function taking the DM pressure into account.
In order to calculate the transfer function in the presence of the DM pressure, we start from the Einstein field equations in the following form,
 \begin{eqnarray}
R_{\mu \nu}=-8\pi G S_{\mu \nu},
 \end{eqnarray}
in which
 \begin{eqnarray}
S_{\mu \nu}=T_{\mu \nu}-\frac{1}{2}g_{\mu \nu}g^{\rho \sigma}T_{\rho \sigma}.
 \end{eqnarray}
The tt-component of this equation leads to \cite{Weinberg},
 \begin{eqnarray}\label{Ts1}
\frac{d}{dt}(a^2 \psi_q)=-4\pi G a^2 (\delta \rho_{DMq}+\delta \rho_{Bq}+2\delta \rho_{Rq}+2\delta \rho_{Nq}+3\delta P_{DMq}).
 \end{eqnarray}
In the above equation, $\psi_q=\frac{\partial}{\partial t}(\frac{h_{ii}}{2a^2})$ and $h_{ii}$ is a small perturbation. Besides, $\delta \rho_{DMq}$, $\delta \rho_{Bq}$, $\delta \rho_{Rq}$, and $\delta \rho_{Nq}$ are the Fourier transformation of density perturbations of dark matter, baryonic matter, radiation, and neutrinos, respectively.
$\delta P_{DMq}$ also shows the Fourier transformation of DM pressure perturbations.
In Eq. (\ref{Ts1}), the EOSs, $P_{Rq}=\frac{1}{3}\rho_{Rq}$ and $P_{Nq}=\frac{1}{3}\rho_{Nq}$ have been applied.

Conservation of energy momentum tensor for the DM sector gives \cite{Weinberg},
 \begin{eqnarray}\label{Ts2}
\delta \dot{\rho}_{DMq}+3H(\delta \rho_{DMq}+\delta P_{DMq})-\frac{q^2}{a^2}(\bar{\rho}_{DM}+\bar{P}_{DM})\delta u_{DMq}= -(\bar{\rho}_{DM}+\bar{P}_{DM})\psi_q,
 \end{eqnarray}
where $\bar{\rho}_{DM}$ and $\bar{P}_{DM}$ are unperturbed density and pressure and $\delta u_{DMq}$ is the Fourier transformation of DM velocity four-vector perturbation. The conservation of energy momentum tensor for the DM sector also results in the following equation,
 \begin{eqnarray}\label{Ts3}
 \delta P_{DMq}+\frac{d}{dt}[(\bar{\rho}_{DM}+\bar{P}_{DM})\delta u_{DMq}]+3H(\bar{\rho}_{DM}+\bar{P}_{DM})\delta u_{DMq}=0.
 \end{eqnarray}
We introduce the parameter $\delta_{\alpha q}$ in the following way,
 \begin{eqnarray}
 \delta_{\alpha q}=\frac{\delta \rho_{\alpha q}}{\bar{\rho}_{\alpha}+\bar{P}_{\alpha}}.
 \end{eqnarray}
Assuming the adiabatic condition, i.e. $ \delta_{B q}= \delta_{R q}$, and using the definition $ \delta_{\alpha q}$,
Eqs. (\ref{Ts1})-(\ref{Ts3}) lead to
 \begin{eqnarray}\label{Ts1A}
\frac{d}{dt}(a^2 \psi_q)= -4\pi G a^2 \{[1+3\frac{\partial P_{DMq}}{\partial \rho_{DMq}}](\bar{\rho}_{DM}+\bar{P}_{DM})\delta_{DM q}+(\bar{\rho}_{B}+\frac{8}{3}\bar{\rho}_{R})\delta_{R q}+\frac{8}{3}\bar{\rho}_{N}\delta_{N q}\},
 \end{eqnarray}
 \begin{eqnarray}\label{Ts2A}
 \dot{\delta}_{DMq}+\{3H(1+\frac{\partial P_{DMq}}{\partial \rho_{DMq}})+\frac{d}{dt}(ln[\bar{\rho}_{DM}+\bar{P}_{DM}])\}\delta_{DMq}-\frac{q^2}{a^2}\delta u_{DMq}= -\psi_q,
 \end{eqnarray}
 \begin{eqnarray}\label{Ts3A}
\frac{\partial P_{DMq}}{\partial \rho_{DMq}} \delta_{DMq}+\{\frac{d}{dt}[ln(\bar{\rho}_{DM}+\bar{P}_{DM})]+3H\}\delta u_{DMq}+
\frac{d}{dt}(\delta u_{DMq})=0.
 \end{eqnarray}
In the matter-dominated era, the terms in Eq. (\ref{Ts1A}) containing $\delta_{R q}$ and $\delta_{N q}$ can be neglected. In the next step, we eliminate $\psi_q$ between Eqs. (\ref{Ts1A}) and (\ref{Ts2A}). The resulting equation is,
 \begin{eqnarray}\label{Tsc1}
 \frac{d}{dt}(f_1\dot{\delta}_{DMq})+\frac{d}{dt}([f_2+f_3]\delta_{DMq})-q^2\frac{d}{dt}(\delta u_{DMq})=f_4\delta_{DMq},
 \end{eqnarray}
with
 \begin{eqnarray}
 f_1=a^2,
 \end{eqnarray}
 \begin{eqnarray}
 f_2=a^2\frac{d}{dt}[ln(\bar{\rho}_{DM}+\bar{P}_{DM})],
 \end{eqnarray}
 \begin{eqnarray}
 f_3=3a^2H(1+\frac{\partial P_{DMq}}{\partial \rho_{DMq}}),
 \end{eqnarray}
 \begin{eqnarray}
 f_4=4\pi G a^2[1+3\frac{\partial P_{DMq}}{\partial \rho_{DMq}}](\bar{\rho}_{DM}+\bar{P}_{DM}).
 \end{eqnarray}
Eq. (\ref{Ts3A}) can also be written as
 \begin{eqnarray}\label{Tsc2}
f_5 \delta_{DMq}+f_6 \delta u_{DMq}+\frac{d}{dt}(\delta u_{DMq})=0.
 \end{eqnarray}
In the above equation,
\begin{eqnarray}
 f_5=\frac{\partial P_{DMq}}{\partial \rho_{DMq}},
 \end{eqnarray}
and
\begin{eqnarray}
 f_6=\frac{d}{dt}[ln(\bar{\rho}_{DM}+\bar{P}_{DM})]+3H.
 \end{eqnarray}
Applying some algebra to eliminate $\delta u_{DMq}$ between Eqs. (\ref{Tsc1}) and (\ref{Tsc2}) results in
\begin{eqnarray}
 \frac{d}{dt}\{\frac{1}{f_6}\frac{d}{dt}(f_1 \dot{\delta}_{DMq})+f_1 \dot{\delta}_{DMq}+\frac{1}{f_6}\frac{d}{dt}([f_2+f_3]\delta_{DMq})\nonumber \\+(q^2 \frac{f_5}{f_6}+f_2+f_3-\frac{f_4}{f_6})\delta_{DMq} \}-f_4 \delta_{DMq}=0.
 \end{eqnarray}
To solve the above equation, the coefficients $f_i$, $i=1-6$, are considered at the present time. Among the three solutions of this equation, only one of them is an increasing function of time. Since two others decay with time, they do not concern us in our calculations. The resulting DM density perturbation in the presence of the DM pressure is substituted into Eq. (\ref{Trans}) to calculate the transfer function for the NZPDM, $T_{NZPDM}$.
This function depends on the dimensionless rescaled wave number
\begin{eqnarray}
k=\frac{19.3(q/a_0)[Mpc^{-1}]}{\Omega_M h^2}.
 \end{eqnarray}
We have shown the transfer function considering NZPDM in Figure \ref{fig6} with the dashed curve. At each wavenumber $k$, the transfer function for the NZPDM is larger than the case of ZPDM. Moreover, the pressure of DM leads to a smaller rate for decreasing of the transfer function with wavenumber.

\vspace{-0.58cm}
\section{Void shapes considering the dark matter pressure}

Starting from the local tidal shear tensor, $T_{ij}$, which is related to the gravitational potential, $\phi$, by
\begin{eqnarray}
T_{ij}=\frac{\partial^2\phi}{\partial x_i \partial x_j}-\frac{1}{3}\nabla^2 \phi \delta_{ij},
 \end{eqnarray}
\cite{Bos}, we study the shapes of the voids considering the DM pressure. The eigenvalues of this tensor are
\begin{eqnarray}
\lambda_1(\mu,\nu)=\frac{1+(\delta_{\upsilon}-2)\nu^2+\mu^2}{\mu^2+\nu^2+1},
 \end{eqnarray}
\begin{eqnarray}
\lambda_2(\mu,\nu)=\frac{1+(\delta_{\upsilon}-2)\mu^2+\nu^2}{\mu^2+\nu^2+1},
 \end{eqnarray}
and $\lambda_3$ which is related to $\lambda_1$ and $\lambda_2$ by $\delta_{\upsilon}=\sum_{i=1}^3\lambda_i$, \cite{Park}.
In the above equations, $\mu$ and $\nu$ are the void's oblateness and sphericity, respectively. In addition, $\delta_{\upsilon}$ is the density contrast threshold for the formation of a void. To study the void shape, the void ellipticity is defined by $\varepsilon\equiv1-\nu$. The probability density distribution for the ellipticity is given by \cite{Lee9,Bos}
\begin{eqnarray}\label{prob}
p(1-\varepsilon;z)&=&p(\nu;z,R_L)=\int_{\nu}^1p[\mu,\nu|\delta=\delta_{\upsilon};\sigma(z,R_L)]d\mu\nonumber \\&=&\int_{\nu}^1d\mu\frac{3375\sqrt{2}}{\sqrt{10\pi}\sigma^5(z,R_L)}exp[-\frac{5\delta_{\upsilon}^2}{2\sigma^2(z,R_L)}
+\frac{15\delta_{\upsilon}(\lambda_1+\lambda_2)}{2\sigma^2(z,R_L)}]\nonumber \\ &\times& exp[-\frac{15(\lambda_1^2+\lambda_1\lambda_2+\lambda_2^2)}{2\sigma^2(z,R_L)}](2\lambda_1+\lambda_2-\delta_{\upsilon})
 (\lambda_1-\lambda_2)\nonumber \\&\times&(\lambda_1+2\lambda_2-\delta_{\upsilon})\frac{4(\delta_{\upsilon}-3)^2\mu\nu}{(\mu^2+\nu^2+1)^3}.
 \end{eqnarray}
In the above equation, $\sigma(z,R_L)$, the linear rms fluctuation
of the matter density field smoothed on a Lagrangian void scale of $R_L$ at redshift, $z$, is as follows \cite{Lee9,Chongchitnan,Bos},
\begin{eqnarray}\label{sig}
\sigma^2(z,R_L)\equiv D^2(z)\int_0^{\infty}\frac{k^2}{2 \pi^2}P_{lin}(k)W^2(kR_L)dk,
 \end{eqnarray}
in which $P_{lin}(k)$ shows the linear matter power spectrum today and $W(kR_L)$, the spherical top-hat function
of radius $R_L$, is as follows,
\begin{eqnarray}
W(kR_L)=3[\frac{sin(kR_L)}{(kR_L)^3}-\frac{cos(kR_L)}{(kR_L)^2}].
 \end{eqnarray}
In the case of ZPDM, the standard linear matter power spectrum is as follows,
\begin{eqnarray}
P_{lin}^{ZPDM}(k)=AkT_{ZPDM}^2(x_{EH}),
 \end{eqnarray}
in which $T_{ZPDM}$ is given by Eq. (\ref{Tzpdm}). The coefficient $A$ in the above equation is determined so that $\sigma(R_L=8h^{-1}\ Mpc)=\sigma_8$ in which $\sigma_8=0.776$. However, for the NZPDM, we apply the following linear matter power spectrum
\begin{eqnarray}
P_{lin}^{NZPDM}(k)=AkT_{NZPDM}^2(k),
 \end{eqnarray}
with the transfer function $T_{NZPDM}$ in the presence of the DM pressure described above.

\begin{figure*}
\vspace*{1cm}       % Give the correct figure height in cm
\includegraphics[width=0.85\textwidth]{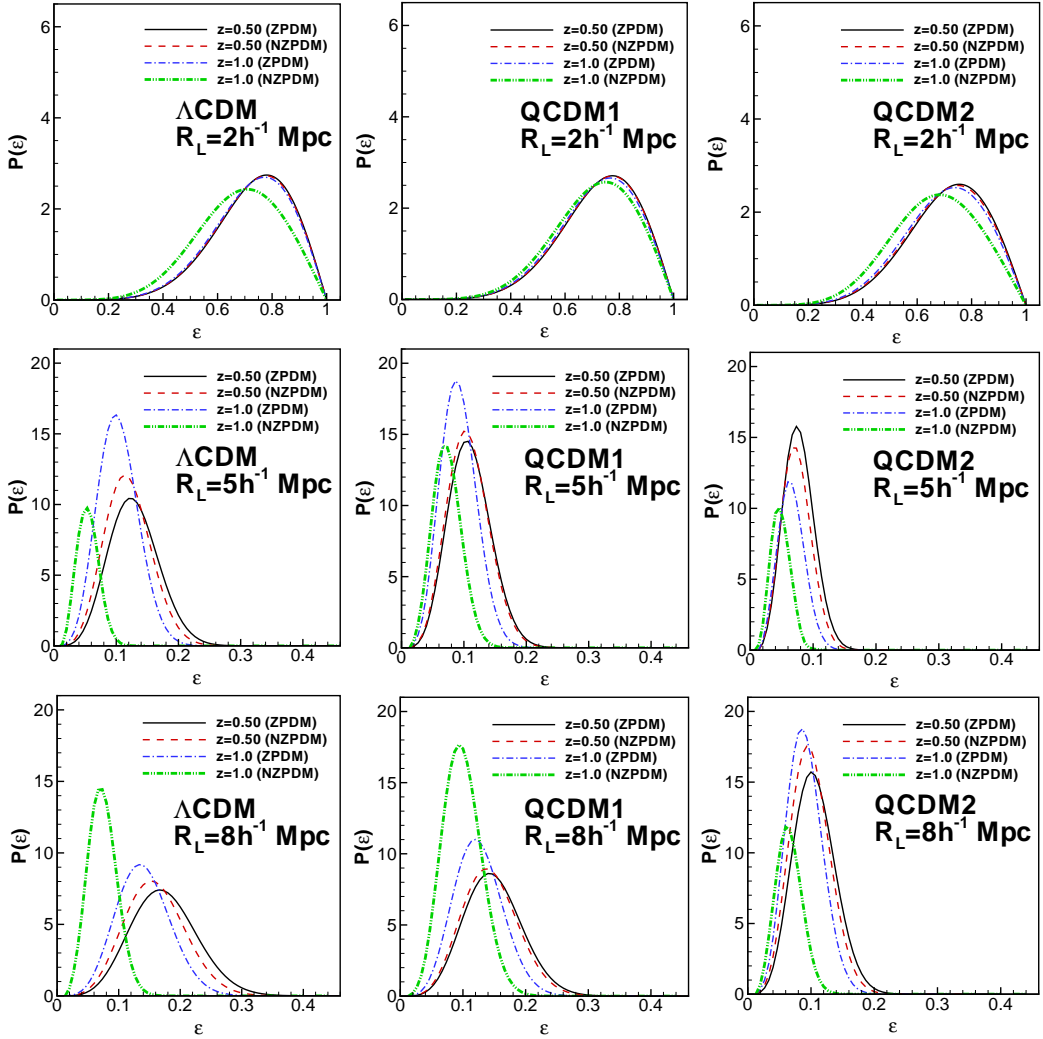}
\caption{Probability density distribution of the void ellipticity at different values of the redshift, $z$, in the cases of zero pressure DM (ZPDM) and non zero pressure DM (NZPDM)
considering different values of the Lagrangian void scale, $R_L$, in different cosmological models.}
\label{fig7}
\end{figure*}

The function $D(z)$ in Eq. (\ref{sig}), is given by Eqs. (\ref{Heath1}) and (\ref{Bonnor12}) for the ZPDM and
NZPDM, respectively.
Figure \ref{fig7} shows the probability density distribution (PDD) for the ellipticity of cosmic voids considering the ZPDM and NZPDM at different values of the redshift, $z$, and various Lagrangian scales, $R_{L}$, in different cosmological models. The void density contrast
has been also set at $\delta_{\upsilon}=-0.9$.
%%%%%%%%%%%%%%%%%%%%%%%%%%%%%%%%%%%%%%%%%%%%%%%%%%%%%%%%%%%%%%%%%%%%%%%%%%%%%%%%%%%%%%%%%%%%%%%%%%%
For the voids with small scales, the effects of DM pressure on the PDD are not considerable at lower values of the redshift. However, the DM pressure affects the PDD if the scale of the voids or the redshift increases.
The DM pressure changes the PDD so that
the voids with smaller ellipticity are more probabilistic.
It means that with the NZPDM, the voids are expected to have more spherical shapes.
Besides, the voids at higher values of the redshift are also more spherical.
Considering the larger scale voids, the PDDs are sharper with their peaks at lower values of the ellipticity.
This is in agreement with the results of \cite{Park} and \cite{Lavaux}.

Each PDD has a maximum value at a specific value of the ellipticity which is denoted by $\varepsilon_{max}$. $\varepsilon_{max}$ determines the value of the ellipticity which the most cosmic voids have.
Table~\ref{table1} presents the values of $\varepsilon_{max}$.
\begin{table*}
\caption{Values of $\varepsilon_{max}$ for the cosmic voids with zero pressure DM (ZPDM) and non zero pressure DM (NZPDM)
at different values of the Lagrangian void scale, $R_L$, applying different cosmological models in the case of
$z=1$.}
\label{table1}       % Give a unique label
\begin{center}  {\footnotesize
\begin{tabular}{|c|c|c|c|c|c|c|}
\hline \multicolumn{1}{|c|}{} & $R_L (h^{-1} Mpc) $  & \multicolumn{1}{c}{} &
\multicolumn{1}{c}{Cosmological Models} &
\multicolumn{1}{c|}{}\\\hline
& & $\Lambda$CDM &QCDM1 &QCDM2    \\\hline
& 2 &0.77   & 0.77    & 0.74      \\
ZPDM &5 &0.10   & 0.09    &  0.06        \\
 &8 &0.13   & 0.12    &0.09
  \\\hline
&2&0.71   &  0.75   &  0.68        \\
 NZPDM&5& 0.05  &  0.07   &  0.05        \\
 &8 & 0.07  &  0.09   &   0.06     \\
\hline
\end{tabular} }
\end{center}
\vspace*{2cm}  % with the correct table height
\end{table*}
%%%%%%%%%%%%%%%%%%%%%%%%%%%%%%%%%%%%%%%%%%%%%%%%%%%%%%%%%%
Table~\ref{table1} confirms that in all cosmological models and at different values of the Lagrangian void scale, the values of $\varepsilon_{max}$ are lower in the case of NZPDM compared to ZPDM.
Therefore, considering the DM pressure, the most cosmic voids are more spherical.
The values of $\varepsilon_{max}$ with ZPDM obtained in QCDM1 model are lower than the $\Lambda$CDM model.
However, considering the case NZPDM, $\varepsilon_{max}$ is higher in QCDM1 model compared to $\Lambda$CDM one. Table~\ref{table1} also verifies that QCDM2 predicts smaller values of $\varepsilon_{max}$ for all voids in both ZPDM and NZPDM cases compared to the other cosmological models.

Applying the PDD, i.e. Eqs. (\ref{prob}), the mean ellipticity of cosmic voids is defined as follows \cite{Lee9},
\begin{eqnarray}
<\varepsilon>=\int_0^1\varepsilon p(\varepsilon;R_L,z) d\varepsilon.
 \end{eqnarray}
 \begin{table*}
\caption{Same as Table~\ref{table1} but for the values of $\langle\varepsilon\rangle$.}
\label{table3}       % Give a unique label
\begin{center}  {\footnotesize
\begin{tabular}{|c|c|c|c|c|c|}
\hline \multicolumn{1}{|c|}{} & $R_L (h^{-1} Mpc) $  & \multicolumn{1}{c}{} &
\multicolumn{1}{c}{Cosmological Models} &
\multicolumn{1}{c|}{}\\\hline
& & $\Lambda$CDM &QCDM1 &QCDM2   \\\hline
& 2 &0.72   & 0.72    &    0.70       \\
ZPDM &5 &0.11   & 0.08    &  0.05         \\
 &8 & 0.15  &  0.14   &0.07
  \\\hline
&2 & 0.68  & 0.71    &    0.66        \\
 NZPDM&5 & 0.05  &   0.05   &  0.05        \\
 &8 & 0.05  &   0.09  &   0.05     \\
\hline
\end{tabular} }
\end{center}
\vspace*{2cm}  % with the correct table height
\end{table*}
Table~\ref{table3} gives the values of the mean ellipticity, $<\varepsilon>$, considering different values of the Lagrangian void scale and applying various cosmological models.
It is obvious that the NZPDM leads to the smaller values of the mean ellipticity compared to ZPDM.
Besides, $<\varepsilon>$ with ZPDM is lower for QCDM1 model compared to $\Lambda$CDM one.
However, for NZPDM, $<\varepsilon>$ in QCDM1 model is higher than the one in $\Lambda$CDM model.
In addition, QCDM2 model gives smaller values of $<\varepsilon>$ compared to other models for all voids in both ZPDM and NZPDM cases.
\begin{figure*}
\vspace*{1cm}       % Give the correct figure height in cm
\includegraphics[width=0.85\textwidth]{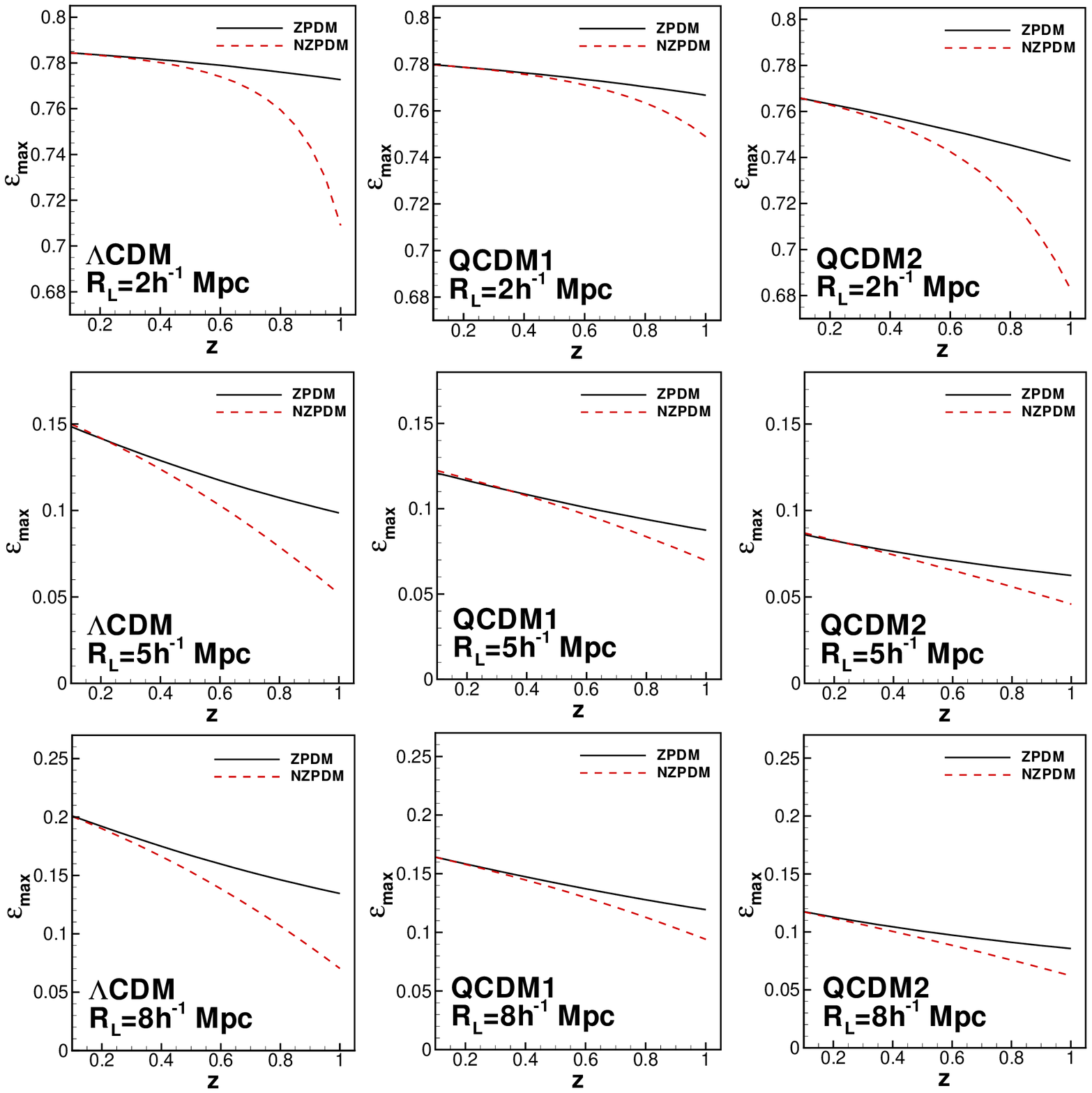}
\caption{Redshift dependency of $\varepsilon_{max}$ for cosmic voids with zero pressure DM (ZPDM) and non zero pressure DM (NZPDM) at different values of the Lagrangian void scale, $R_L$, applying different cosmological models.}
\label{fig8}
\end{figure*}

The redshift dependency of the quantity $\varepsilon_{max}$ has been shown in Figure \ref{fig8}.
In all cosmological models and considering the voids with different scales, $\varepsilon_{max}$ reduces with the redshift for both ZPDM and NZPDM cases.
However, the rate at which $\varepsilon_{max}$ decreases with the redshift is higher in voids with NZPDM.
At each redshift, $\varepsilon_{max}$ is lower for the voids with NZPDM.
The NZPDM influences $\varepsilon_{max}$ more considerably at higher values of the redshift.
The effects of NZPDM on the redshift dependency of $\varepsilon_{max}$ are more considerable in $\Lambda$CDM model.
The values of $\varepsilon_{max}$ versus the redshift depend on the cosmological models which are higher in $\Lambda$CDM, QCDM1, and QCDM2, respectively.

\begin{figure*}
\vspace*{1cm}       % Give the correct figure height in cm
\includegraphics[width=0.85\textwidth]{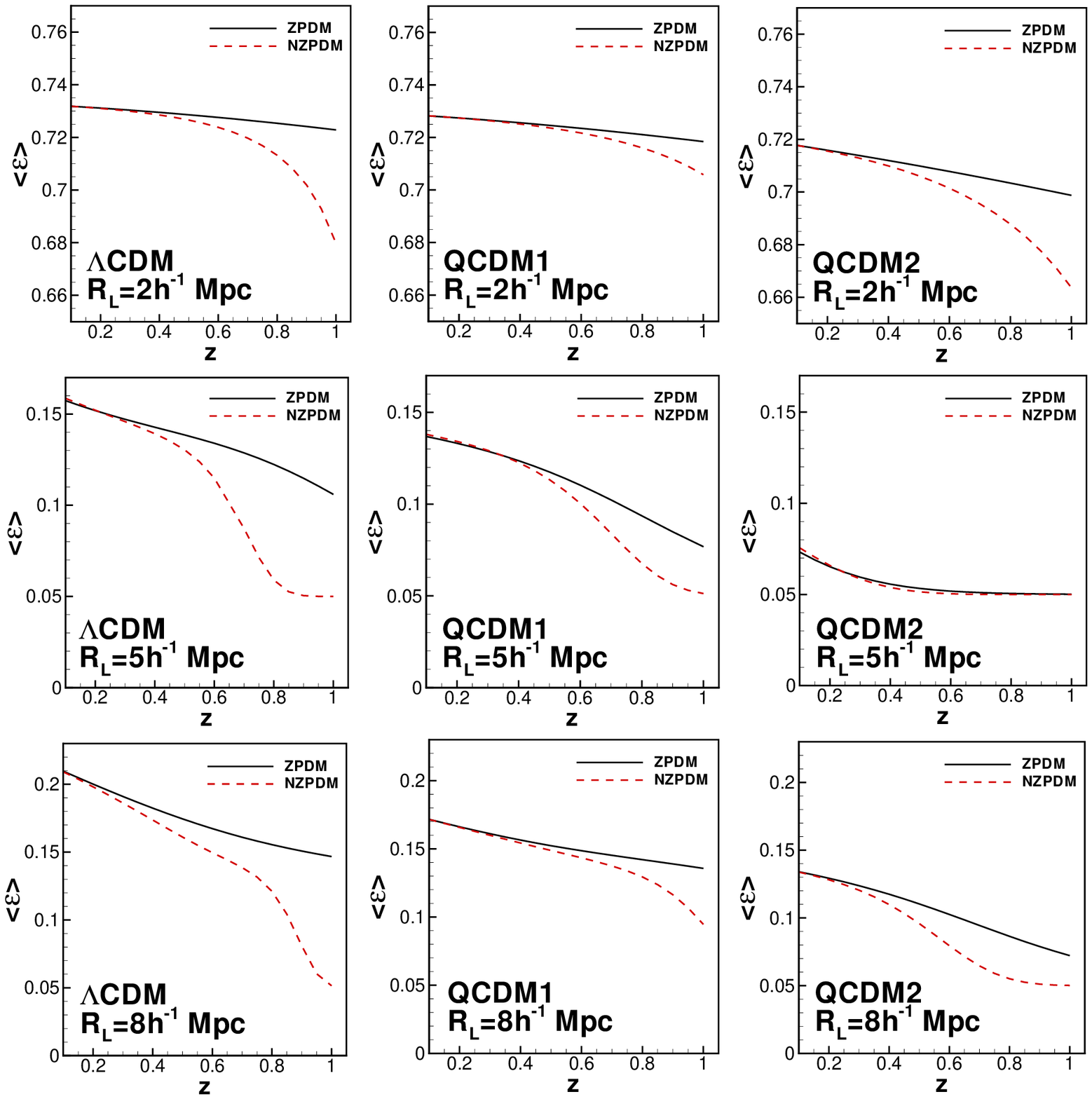}
\caption{Same as Figure \ref{fig8} but for the values of $<\varepsilon>$.}
\label{fig10}
\end{figure*}

Figure \ref{fig10} gives the values of the mean ellipticity, $<\varepsilon>$, as a function of the redshift for cosmic voids with ZPDM and NZPDM.  The mean ellipticity decreases as the redshift grows, similar to the results of \cite{Lee9,Bos,Ricciardelli}.
This reduction is more significant when the pressure of the DM is considered.
$<\varepsilon>$ is lower in the cases of NZPDM compared to ZPDM ones.
Similar to $\varepsilon_{max}$, the effects of the DM pressure is more considerable in $\Lambda$CDM model.
%%%%%%%%%%%%%%%%%%%%%%%%%%%%%%%%%%%%%%%%%%%%%%%%%%%%%%%%%%%%%%%%%%%%%%%%%%%%%%%%%%%%%%%%%%

\vspace{-0.58cm}
\section{Summary and Conclusions}
\label{sec:3}
The dark matter equation of state from the pseudo-isothermal density profile of galaxies
has been employed to study the shapes of cosmic voids in different cosmological models.
Our results confirm that for the cosmic voids at higher values of the redshift, the dark matter pressure alters the probability density distribution for the ellipticity of cosmic voids.
Considering the dark matter pressure, the voids with smaller ellipticity are more probabilistic. In other words, the voids with dark matter pressure are expected to have more spherical shapes.
We have also shown that the mean ellipticity of cosmic voids is smaller when the pressure of dark matter is considered. In addition, the rate at which the mean ellipticity decreases with the redshift is affected by the dark matter pressure.

\section*{Acknowledgements}
The author wishes to thank the Shiraz University Research Council.

\end{document}